# Aftershocks of the 2013 deep Okhotsk earthquake

Anatol V. Guglielmi, Oleg D. Zotov

*Institute of Physics of the Earth RAS, Moscow, Russia*

[guglielmi@mail.ru](mailto:guglielmi@mail.ru), [ozotov@inbox.ru](mailto:ozotov@inbox.ru)

**Abstract**

Strong deep earthquakes occurred in the Sea of Okhotsk, May 24, 2013 at 05:44 UTC. The magnitude of the earthquake is M = 8.3, the depth of the hypocenter is 629 km. The earthquake was accompanied by the deep-focus aftershocks. The paper is devoted to the study of the space-time properties of aftershocks. An estimate is made of the deactivation coefficient of the earthquake source, which "cools down" after the main shock. The modulation of aftershocks by toroidal oscillations of the Earth was discovered. The questions about the round-the-world seismic echo and the ribbed structure of the spatio-temporal distribution of epicenters are considered.

*Keywords*: Omori law, deactivation coefficient, inverse problem, modulation, free oscillations, round-the-world echo, space-time distribution

**Contents**

## 1. Introduction

Omori's law [1] states that the number of aftershocks $n(t)$ in the epicentral zone of a strong earthquake decreases hyperbiolically over time:

$$n(t) = \frac{k}{c+t}. \tag{1}$$

It is assumed here that the parameters $k$ and $c$ are independent of time.

As we know, the hyperbolic Omori's law was the first law of the physics of earthquakes [2]. Recently, a generalization of the Omori law was proposed [3], which makes it possible to take into account non-stationarity of geological environment in the earthquake source:

$$n(t) = \frac{n_0}{1 + n_0 \int_0^t \sigma(t')dt'}. \tag{2}$$

Here $\sigma(t)$ is the coefficient of deactivation of the earthquake source, "cooling down" after the main shock. We draw attention to the fact that up to notation formula (2) coincides with (1) if $\sigma = \text{const}$.

Equation (2) makes it possible to formulate and solve the inverse problem of an earthquake source [2]. The essence of the inverse problem is that one should find the deactivation coefficient $\sigma(t)$ as a function of time from the observation data of the aftershock frequency $n(t)$.

Previously, the evolution of $\sigma(t)$ was studied using data from aftershocks after strong earthquakes, the hypocenters of which were located at a relatively shallow depth [4–6]. In contrast to this, in this paper we present the result of solving the inverse problem for the deep-focus Sea of Okhotsk earthquake. In addition, we will consider the Gutenberg-Richter and Bath laws, and also consider a number of interesting properties of the deep aftershocks.

## 2. Deactivation coefficient

The strong deep-focus earthquake occurred in the Sea of Okhotsk on May 24, 2013 at 05:44 UTC. The magnitude of the earthquake is M = 8.3, the depth of the hypocenter is 629 km. The coordinates of the epicenter: 54.86° N, 153.41° E [7–10].

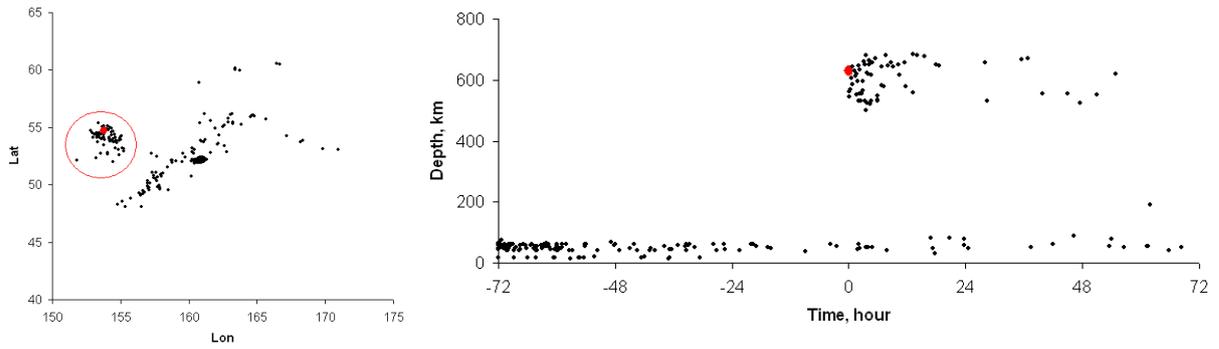

Fig. 1. Epicenter (left) and hypocenter (right) of the main shock of the Sea of Okhotsk M8.3 earthquake are marked with red dots. The red circle outlines the deep-focus aftershocks.

Figure 1 gives an idea of the location of the main shock (red dot) and associated aftershocks. To construct the figure we used the data from the Earthquakes Catalogue for Kamchatka and the Commander Islands (http://sdis.emsd.ru/info/earthquakes/catalogue.php). The positions of the epicenters in the left figure are plotted according to the registration data from May 1 to May 31, 2013. We will focus on the deep-focus events. The epicenters of these events are circled in red in the left picture. In the right picture we see that deep-focus aftershocks form a compact group at depths of 500–680 km.

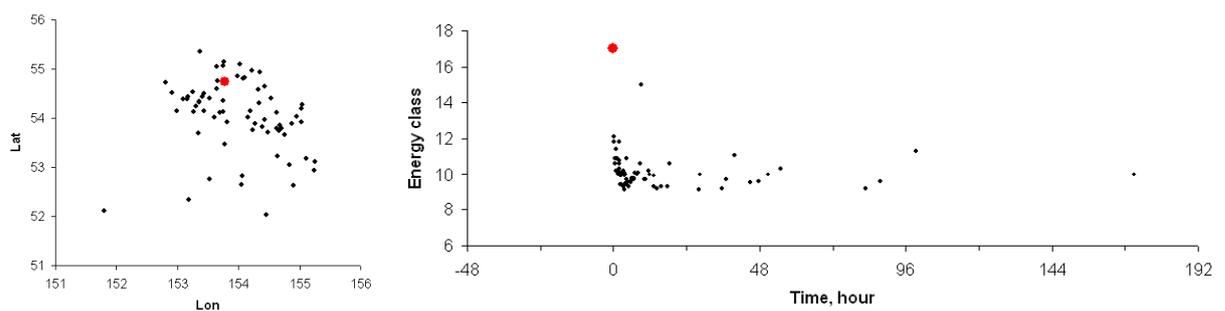

Fig. 2. Epicenters (left) and energy class (right) of deep-focus earthquakes.

On the left panel of Figure 2, we see the location of the epicenters in a larger plan. The right panel gives an idea of the energetic class of deep-focus aftershocks. For reference, here is a formula linking local magnitude with energy class: $M_L = 0.5K - 0.75$.

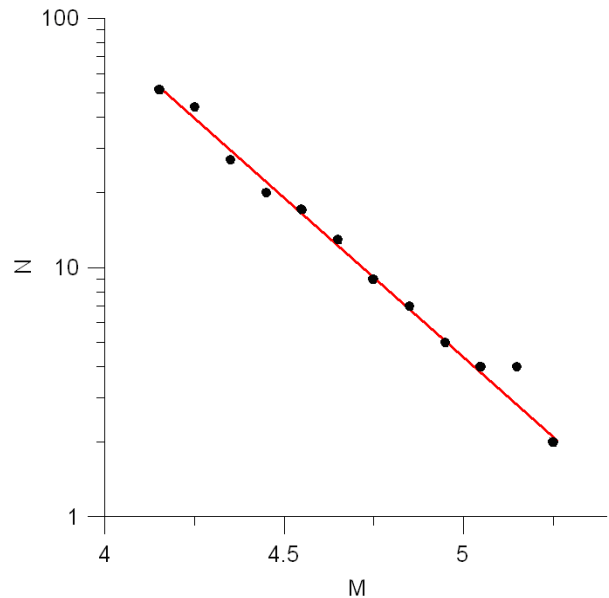

Fig. 3. Gutenberg-Richter distribution for the deep-focus aftershocks.

The Gutenberg-Richter law [11]

$$\lg N = a - bM \tag{3}$$

holds for deep aftershocks, as seen in Figure 3. Here $N$ is the number of aftershocks with magnitude $M$ and above, $a = 7.03$, $b = 1.3$.

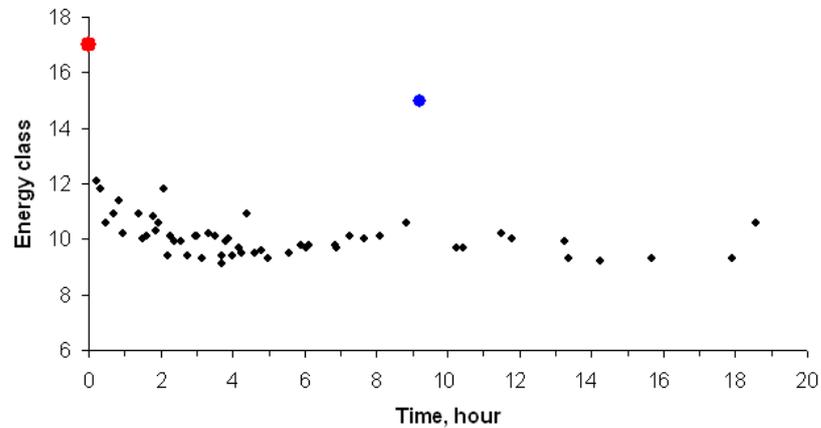

Fig. 4. The main shock (red dot) and the strongest aftershock (blue dot).

The Law of Bath [12] is also being performed. In Figure 4, the red dot denotes the main shock. The strongest aftershock is marked with a blue dot. It occurred after 9 hours 11 minutes 42 seconds after the main shock. The difference between the local magnitudes of the main shock and the strongest aftershocks is $\Delta M_L = 1$.

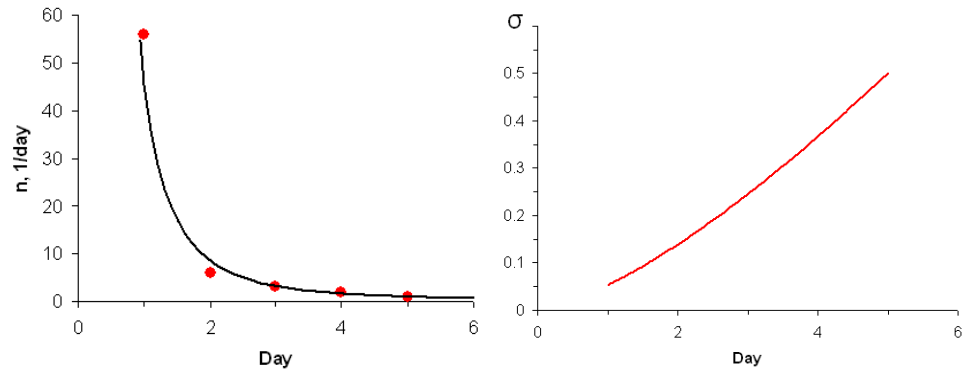

Fig. 5. Frequency of deep aftershocks (left) and deactivation coefficient of the earthquake source (right).

On the left in Figure 5, the dependence of the frequency of aftershocks on time is shown, and on the right, the solution of the inverse problem of the source: $\sigma = 0.053 \cdot t^{1.4}$. Let us note two features of the evolution of deep-focus aftershocks:

1. Abnormally high deactivation coefficient.
2. The absence of the so-called "Omori epoch", i.e. the time interval in which $\sigma = \text{const}$ as it usually happens in the evolution of shallow aftershocks (see. [4–6]).

## 3. Discussion
### 3.1. Aftershock spectroscopy

The source of the earthquake is affected by endogenous and exogenous triggers, leading to a deviation from the Omori law. One of the endogenous triggers is the free oscillations of the Earth. The action of this trigger leads to modulation of shallow-focus aftershocks at the frequencies of spheroidal and toroidal oscillations [13–17].

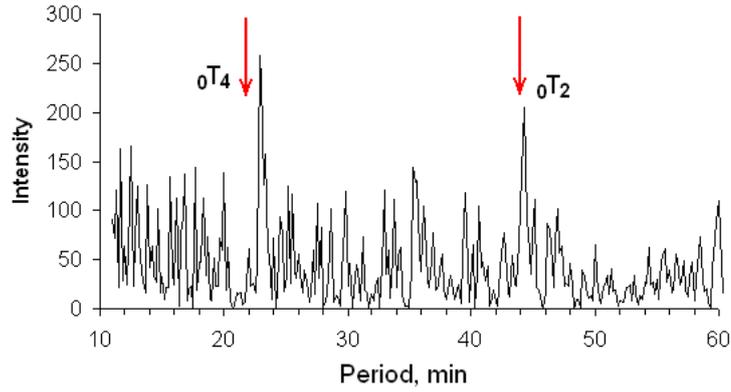

Fig. 6. Aftershock flux spectrum. The arrows indicate the periods of the Earth's toroidal oscillations.

The periodogram in Figure 6 shows that the deep source apparently also exposed to the influence of the free oscillations of the Earth. We see two peaks with the periods close to the periods of two modes of toroidal oscillations $_0T_2 = 43.94$ min, and $_0T_4 = 21.72$ min (see for example [18]). This issue certainly deserves further study.

### 3.2. About the round-the-world seismic echo

In addition to the periodic trigger, which was discussed above, there is also a pulse trigger. It originates at the main impact in the form of a circular surface wave. About 3 hours later the wave having circled the globe returns to the epicenter of the main shock. The phenomenon was called by us a round-the-world seismic echo. From geometric considerations it is clear that the intensity of seismic vibrations increases monotonically as the echo approaches the epicenter. As a result, the round-the-world echo can induce powerful aftershocks.

The analysis unconditionally confirmed this prediction in the case of strong earthquakes at a relatively shallow depth [16, 19, 20]. Generally speaking, the deep-focus earthquakes are much less effectively excite surface waves. Probably for this reason, we did not find clear signs of intensification of deep aftershocks 3 hours after the Sea of Okhotsk earthquake.

## 3.3. Ribbed structure of aftershocks

In a series of studies [21–24] devoted to the space-time dynamics of shallow-focus aftershocks a highly unusual structure of event distribution on the x-t plane was discovered. Here x is the epicentral distance, t is the time elapsed after the main shock. At a quick glance, the distribution pattern resembles the Rorschach Ink-blot. It is known that the validity of the Rorschach test sometimes raises doubts even in psychoanalysis. However, careful checks have shown that the ribbed (corrugated) structure of the spatio-temporal distribution of aftershocks does indeed appear on the x-t plane.

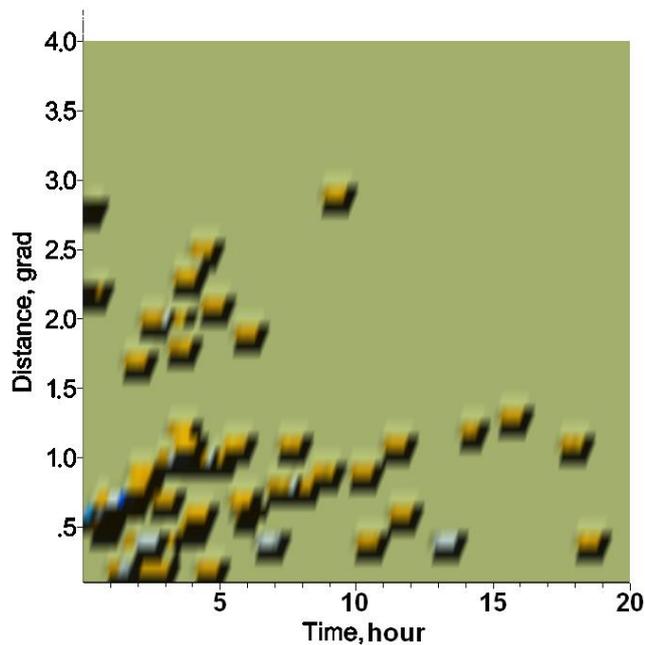

Fig. 7. Spatial-temporal distribution of aftershocks.

We tried to find signs of a ribbed structure in the distribution of the Sea of Okhotsk aftershocks. The result is shown in Figure 7. The picture turned out to be not as distinct as in the case of shallow aftershocks. Nevertheless, it seems to us that one can see elements of a ribbed structure similar to that found for shallow-focus aftershocks.

We have the impression that some diffusion process unknown to us generates the observed pattern. Therefore, we associate a certain hope of reaching an understanding with the search for nontrivial solutions of the well-known Kolmogorov-Fisher equation

$$\frac{\partial n}{\partial t} + \sigma n^2 = D\nabla^2 n. \tag{4}$$

Here $\nabla$ is the 2D Hamilton operator, $D$ is the phenomenological coefficient of diffusion.

## 4. Conclusion

The result of a preliminary analysis of the Otomomorsk earthquake shows that this outstanding event deserves further deeper investigation. We have only touched the Omori law and briefly reviewed the interesting properties of deep aftershocks.

In conclusion, we would like to highlight the issue of the role of laws in the physics of earthquakes. It is quite clear that the search for laws is more important than a simple selection of formulas for approximating observation data. When studying earthquakes, Omori's law, Gutenberg-Richter's law, Bath's law, the laws of the spatial distribution of aftershocks and others are used. To be specific, let's focus on the Omori Law.

The hyperbolicity of the evolution of aftershocks in the form (1) contradicts observations. Under the influence of this circumstance, Hirano [25] and Utsu [26–28] replace the one-parameter Omori formula (1) with the two-parameter fitting formula $n(t) = k/(c+t)^p$. Here parameter $p$ changes from one event to another.

In contrast, we prefer to keep the simple and, in our own way, beautiful idea of hyperbolicity. To eliminate the contradiction between observations and the law in the form (1), we introduced the concept of the source deactivation coefficient and replaced the Omori hyperbola (1) with the shortened Bernoulli equation [2–6]:

$$\frac{dn}{dt} + \sigma n^2 = 0. \tag{5}$$

This is a one-parameter equation. It is completely equivalent to the evolution equation (2), which clearly demonstrates the hyperbolic structure of the law.

We prefer to keep the hyperbolic law, and here's why: we have an interesting historical analogy. At one time it was found that, generally speaking, the planets do not move along ellipses. Nevertheless, astronomers preferred to preserve the quadratic law of gravity and made efforts to find the reason for the mismatch in some incidental factors (the influence of planets on each other, the finite velocity of light propagation). The opposite approach, namely the rejection of the law, would deprive us of all support. Following the law led to the discovery of unknown planet, to the measurement of the speed of light from the data on the motion of Jupiter's moons, and so on.

Sometimes we hear that it is wrong to compare the Law of Universal Gravitation and the Omori Law. One can object to this as follows. In any case it is worth learning at least some lesson from history. Of course, Newton's and Omori's laws are incomparable in significance and universality. But a distant parallel can be traced from the point of view of the psychology of the search for new laws.

Our conclusion from the above reasoning is that at the beginning of the last century it was not worth abandoning the idea of hyperbolicity until the possibilities of its development were fully used.

The expanded version of this paper will be published in the Journal of Volcanology and Seismology.

*Acknowledgments*. We express our gratitude to A.D. Zavyalov and B.I. Klain for numerous discussions of earthquake physics at the joint implementation of the experimental and theoretical study of aftershocks. We also thank A.S. Potapov for his interest in our work and valuable advice. The work was supported by the project of the RFBR 18-05-00096 and the state assignment program of the IPhE RAS.